\documentclass[aps,prb,a4paper,preprint,showpacs]{revtex4}
\usepackage{amsmath}
\usepackage{amsfonts}
\usepackage{amssymb}
\usepackage{slashbox}
\usepackage{CJK}
\usepackage{graphicx}
\usepackage{amsbsy}

\begin{document}
\title{Spin-dependent pump current and noise in an adiabatic
quantum pump based on domain walls in a magnetic nanowire}
\author{Rui Zhu$^{1}$\renewcommand{\thefootnote}{*}\footnote{Corresponding author. Electronic address:
rzhu@scut.edu.cn} and Jamal Berakdar$^{2}$  }
\address{$^{1}$Department of Physics, South China University of Technology,
Guangzhou 510641, People's Republic of China \\ $^{2}$ Institut
f\"{u}r Physik, Martin-Luther-Universit\"{a}t Halle-Wittenberg,
D-06120 Halle, Germany}

\begin{abstract}
We study the pump current and noise properties in an
adiabatically modulated magnetic nanowire  with double domain walls
(DW). The modulation is brought about by applying a slowly
oscillating magnetic and electric fields with a controllable phase
difference. The pumping mechanism  resembles  the
case of the quantum dot pump with two-oscillating gates. The pump current,
shot noise, and heat flow show peaks when the Fermi energy matches
with the spin-split resonant levels localized between the DWs. The
peak height of the pump current is an indicator for the lifetime of
the spin-split quasistationary states between the DWs. For sharp
DWs, the energy absorption from the oscillating fields results in
 side-band formations observable  in the pump
current. The pump noise carries information on the  correlation
properties between the nonequilibrium electrons and the quasi-holes
created by the oscillating scatterer. The ratio between the pump
shot noise and the heat flow serves as an indicator for
quasi-particle correlation.

\end{abstract}


\maketitle

\narrowtext

\section{Introduction}

Since the first experimental realization of the quantum
pump\cite{Ref1}, research on quantum charge and spin pumping has
attracted increasing interest
\cite{Ref2,Ref3,Ref4,Ref5,Ref6,Ref7,Ref8,Ref9,Ref10,Ref11,Ref12,Ref13,Ref14,Ref15,Ref16,Ref17,Ref18,Ref19}.
The current and noise properties in various quantum pump structures
and devices were investigated such as magnetic-barrier-modulated two
dimensional electron gas\cite{Ref4}, mesoscopic one-dimensional
wire\cite{Ref6}, quantum-dot structures\cite{Ref5,Ref11,Ref12},
 mesoscopic rings with Aharonov-Casher and Aharonov-Bohm
effect\cite{Ref7}, magnetic tunnel junctions\cite{Ref10}.
Correspondingly, theoretical techniques have been put forward for
the treatment of
 the quantum pumps (Refs.(\onlinecite{Ref2,Ref3,Ref18}) and references therein).
 Of particular interest for the present work is
 the scattering matrix approach for ac transport,  as detailed  by
 Moskalets \emph{et al}.\cite{Ref3}  who derived
 general expressions for the pump current, heat flow, and the
shot noise for an adiabatically driven  quantum pumps in the weak
pumping limit. The pump current was found to vary in a sinusoidal
manner as a function of the phase difference between the two
oscillating potentials. It increases linearly with the frequency in
line  with the experimental finding. Recently, Park \emph{et
al}.\cite{Ref5} obtained an expression for the admittance and the
current noise for a driven nanocapacitor in terms of the Floquet
scattering matrix and derived a nonequilibrium
fluctuation-dissipation relation. The effect of weak
electron-electron interaction on the noise was investigated by
Devillard \emph{et al}.\cite{Ref6} using the scattering matrix
renormalized by interactions. Applying the Green's function
approach, Wang \emph{et al}.\cite{Ref14,Ref15,Ref16} presented a
nonperturbative theory for the parametric quantum pump at arbitrary
frequencies and pumping strengths. Independently,
Arrachea\cite{Ref17} presented a general treatment based on
nonequilibrium Green functions to study transport phenomena in
quantum pumps.

The shot noise properties of a quantum pump are important in two
aspects: Understanding the underlying mechanisms of the shot noise
may offer possible ways to improve pumping efficiency and achieve
optimal pumping\cite{Ref4,Ref6,Ref10,Ref12}. On the  other hand  the
shot noise reflects current correlation and is sensitive to the pump
source configuration.
 For transport in mesoscopic systems
 it is shown that shot noise carries
 information beyond those obtainable from
conductance measurements\cite{Ref20,Ref21,Ref22,Ref23} such as
quantum correlation of electrons\cite{Ref21} including spin-orbit
coupling effect\cite{Ref22}, and entanglement\cite{Ref23}.

 In this work, we focus on the current and shot noise
properties in a particular spin-dependent quantum pump  based on two
domain walls in a magnetic quantum wire (shown in Fig.1). In general, the  transport
properties of magnetic domain walls (DWs) are actively discussed and
realized for
 spintronics applications (cf. Ref.\onlinecite{Ref24,Ref25,Ref26,Ref27,Ref28} and references therein).
  To our knowledge however, a DW-based
 quantum pump has not yet been considered.
As shown by Dugaev \emph{et al}.\cite{Ref24} in a semiconducting
magnetic nanowire,
  double DWs separated by a distance less than the phase coherence length
 act as a spin quantum well within which quasistationary spin-dependent quantized states
 are formed. To  generate in this structure a spin-polarized dc current at zero
 bias voltage we propose to apply a slowly
oscillating gate potential and a varying magnetic field. In effect
this means a periodic  variation of the chemical potential and the
scattering strength of the DWs. In the spirit of an adiabatic
quantum pump, these varying parameters should oscillate slowly
relative to the carriers' interaction time with the DWs. The DWs
themselves however should be sharp and not adiabatic. This renders a
strong scattering from DWs and hence the formation of the quantum
well. Here we note that conventionally a DW is called adiabatic when
its extensions is larger than the Fermi wave length of the carriers
\cite{Ref26}. Therefore, for an experimental realization low-density
magnetic semiconductor based nanowires are favorable, e.g. as
 reported in Ref.\onlinecite{Ref25}. In
the following, we investigate the pumping current and shot noise
characteristics of this system and explore the pumping properties
and the underlying relation between the pump noise and quantum
correlation.
\section{Theoretical formulation}
As shown schematically in Fig.(1),
we consider a magnetic nanowire with a magnetization profile
consisting of two DWs separated by the distance $2L$.
 The phase coherence length
is larger than $2L$. The width of
each  of the DWs is $2 \delta$. The magnetization vector field
${\bf{M}}\left( z \right)$ in both DWs varies within the $x$-$z$
plane with the $z$ axis being along the wire. Thus, $z$ is
the easy axis, and the $x$-$z$ plane is the easy plane.
As illustrated in Fig.1 we assume further that the magnitude
of  ${\bf{M}}\left( z \right)$  is hardly changed but its direction, i.e.
we can write
${\bf{M}}\left( z \right)=M{\bf{n}}\left( z \right)$,
where ${\bf{n}}\left( z \right)$
is a unit vector field.
We study the
case where the thickness and the width of the wire are small such
that only one size-quantized level (i.e. only a single transverse
subband) is populated. Such a system is achievable  for  magnetic
semiconductor-based structures with an appropriately tuned
 carrier density. Adopting a continuum model,
 we describe the
independent carriers motion along the wire coupled to the
non-collinear  magnetization field $ {\bf{M}}(z)$
with a strength determined by the Kondo-type  coupling constant
 $J$.   The single-particle Hamiltonian reads then \cite{Ref26}
\begin{equation}
H =  - \frac{{\hbar ^2 }}{{2m}}\frac{{d^2 }}{{dz^2 }} - \Theta
{{n}}_z\left( z \right) \sigma _z  - \Theta {{n}_x}\left( z \right)
\sigma _x,
\label{eq:h} \end{equation}
where $\Theta=JM$ and  $n_{x(z)}$ is the
$x(z)$ component of ${\bf{n}}\left( z \right)$, and $m$
is the carrier's effective mass. $\Theta$ is tunable externally, e.g. by a magnetic field.
 The potential profile is shown in Fig.1.
  The  carriers' wave functions  are expressible as
\begin{equation}
 \begin{array}{l}
 \begin{array}{*{20}c}
   {\psi _k \left( z \right) = \left[\left( {e^{ikz}  + Re^{ - ikz} } \right)\left|  \uparrow  \right\rangle  + R_f e^{\kappa z} \left|  \downarrow  \right\rangle\right] e^{-iEt/\hbar},} & {z <  - L,}  \\
\end{array} \\
 \begin{array}{*{20}c}
   {\psi _k \left( z \right) = \left[\left( {Ae^{\kappa z}  + Be^{ - \kappa z} } \right)\left|  \uparrow  \right\rangle  + \left( {Ce^{ikz}  + De^{ - ikz} } \right)\left|  \downarrow  \right\rangle \right] e^{-iEt/\hbar},} & {\left| z \right| < L,}  \\
\end{array} \\
 \begin{array}{*{20}c}
   {\psi _k \left( z \right) = \left[Te^{ikz} \left|  \uparrow  \right\rangle  + T_f e^{ - \kappa z} \left|  \downarrow  \right\rangle \right] e^{-iEt/\hbar},} & {z > L,}  \\
\end{array} \\
 \end{array} %\times e^{{{ - iEt} \mathord{\left/
% {\vphantom {{ - iEt} \hbar }} \right.
%\kern-\nulldelimiterspace} \hbar }} ,
\end{equation}
where $ \left|  \uparrow  \right\rangle$ ($ \left|  \downarrow
\right\rangle$) is the spin-up (spin-down) component of the carrier
states, $ k = [2m(E  + JM)]^{1/2} /\hbar $, and $ \kappa = [2m(JM -
E )]^{1/2} /\hbar $. As illustrated in Fig.1, we measure the electron energy $E$
 from the
midpoint between spin-up and spin-down band edges. The chemical potential $\mu$
is set by the particle density and can be tuned by an external gate $X_1$.
The non-spin-flip
(spin-flip) transmission and reflection coefficients $T$ and $R$
($T_f$ and $R_f$) as well as the constants $A$, $B$, $C$, and $D$
can be deduced from the solutions of Eq. (1) and from the wave function
continuity requirements. We note, that instead of using the wave function
we can equivalently utilize the transmission and reflection amplitudes \cite{kidun}.
For the system depicted in Fig.1 the transmission and the reflection amplitudes
$T(\Theta,\mu)$ and $R(\Theta,\mu)$  were
derived and given explicitly in Ref.\cite{Ref24} (same applies to ($T_f$ and $R_f$)).

Following the standard scattering approach\cite{Ref2, Ref3, Ref29}
we introduce the fermionic creation and annihilation operators for
the carrier scattering states. The operator $ \hat a_{L\sigma
}^\dag (E) $ or $ \hat a_{L\sigma } (E) $ creates or annihilates
particles with total energy $E$ and spin polarization $\sigma$ in
the left lead, which are incident upon the sample. Analogously, we
define the creation $ \hat b_{L\sigma }^\dag (E) $ and annihilation
$ \hat b_{L\sigma } (E) $ operators for the  outgoing
single-particle states.  For magnetic semiconductor nanowires with a
moderate carrier density the chemical potential $\mu $ is tunable to
be in one of the magnetically split subbands. In this case one
achieves a full spin polarization of the electron gas.
 Therefore,
the incident electrons are fully spin polarized.
% Considering a
% particular incident energy $E$, the problem reduces to
%  a one-channel scattering
%problem.
We note that the spin-down part of
the wave function decays outside the double-domain-wall (DW) regime (cf. Fig.1).
 Correspondingly, only
spin-up electrons tunnel through the barrier and contribute to the
conductance.
% The regular two spin channels reduce to one (spin-up
%one) in this case.
Furthermore, the energy is conserved during the tunneling
process \cite{Ref30}. As mentioned above, the wire is such that
a single
one-dimensional subband is populated and hence
 the transverse channels are not considered here.
The scattering matrix $S$ is  follows from the relation
\begin{equation}
\left( {\begin{array}{*{20}c}
   {b_{L \uparrow } }  \\
   {b_{R \uparrow } }  \\
\end{array}} \right) = \underbrace {\left( {\begin{array}{*{20}c}
   R & {T'}  \\
   T & {R'}  \\
\end{array}} \right)}_S\left( {\begin{array}{*{20}c}
   {a_{L \uparrow } }  \\
   {a_{R \uparrow } }  \\
\end{array}} \right),
\end{equation}
where, as a result of the structure configuration symmetry the
relation $ T' = T$, $R' = R $ applies\cite{Ref31}. States that decay
exponentially away from the DWs    do not contribute directly to the
current flow and thus to the scattering matrix. We note that the
external magnetic field will be  incorporated as an induced change
of $\Theta$ in Eq.(\ref{eq:h}), i.e. as an effective change in the
height of DWs.

%

%Here,
%the incident and outgoing electrons have the same momentum and
%effective mass.
%Thus, the scattering matrix ($S$ matrix) defined in
%terms of the current amplitude which is equal to the wave amplitude
%times the square root of the velocity is the same as that expressed
%in terms of the wave amplitude in Eq. (5).

In the adiabatic regime the external perturbations vary  slowly on the scale
of the carriers interaction time (Wigner delay time) with the DWs structures.
In this case one can employ an
instant scattering matrix approach, i.e. $S(t)$ depends only parameterically
on the time $t$.
To realize a quantum  pump one varies simultaneously  two system
parameters, e.g. \cite{Ref2,Ref3}
\begin{equation}
\begin{array}{l}
 X_1 \left( t \right) = X_{10}  + X_{\omega ,1} e^{i\left( {\omega t - \varphi _1 } \right)}  + X_{\omega ,1} e^{ - i\left( {\omega t - \varphi _1 } \right)} , \\
 X_2 \left( t \right) = X_{20}  + X_{\omega ,2} e^{i\left( {\omega t - \varphi _2 } \right)}  + X_{\omega ,2} e^{ - i\left( {\omega t - \varphi _2 } \right)} . \\
 \end{array}
 \end{equation}
Here, $X_1$ is a measure for the carrier coupling energy to the DWs (cf. eq.(1))
$\Theta  \equiv JM$ which can be modulated
 by applying a low-frequency ($\omega$)
alternating external magnetic field.  $X_2$ is the Fermi level
position $\mu $, which is adiabatically varied by exposing the
device to ac gate potential. $X_{\omega ,1} $ and $X_{\omega ,2} $ are the
corresponding oscillating amplitudes with   phases $\varphi_{1/2}$;
  $X_{10}$ and $X_{20}$ are the static (equilibrium) components.

Different from the widely discussed adiabatic spin pump based on an
normal metal/ferromagnet/normal metal junction, in our considered system, the direction of the
magnetization in the nanowire and DWs remain constant in time: it
is along $z$ direction outside the double DW regime and along $-z$
direction between the two DWs; the magnetization of the DWs varies
 in the $x$-$z$ plane (cf. Fig. 1).  Applying an alternating
external magnetic field, the strength of the magnetization
varies while out-of-plane precession is suppressed by magnetic anisotropy.
Hence, effects related to field-induced magnetization precession
\cite{Ref18, Ref19} are not considered here. For the reason that the scattering
process has only one spin channel (spin-up channel), the pumped
current is spin polarized, as detailed below.

   As in the work of  Moskalets and B\"uttiker \cite{Ref3},
 in the weak pumping limit ($X_{\omega ,j}  \ll
X_{j0} $) and at zero temperature, the spin-polarized pump current,
noise, and heat flow could be expressed in terms of the scattering
matrix as follows.
\begin{equation} I_\alpha   =
\frac{{e\omega }}{{2\pi }}\sum\limits_{\beta j_1 j_2 } {X_{\omega
,j_1 } X_{\omega ,j_2 } \frac{{\partial S_{\alpha \beta }^*
}}{{\partial X_{j_1 } }}\frac{{\partial S_{\alpha \beta }
}}{{\partial X_{j_2 } }}2\sin \left( {\varphi _{j_1 }  - \varphi
_{j_2 } } \right)},
\end{equation}
\begin{equation}
\hspace{0.1cm} H_\alpha   = \frac{{\hbar \omega ^2 }}{{4\pi
}}\sum\limits_{\beta j_1 j_2 } {X_{\omega ,j_1 } X_{\omega ,j_2 }
\frac{{\partial S_{\alpha \beta } }}{{\partial X_{j_1 }
}}\frac{{\partial S_{\alpha \beta }^* }}{{\partial X_{j_2 } }}2\cos
\left( {\varphi _{j_1 }  - \varphi _{j_2 } } \right)},
\end{equation}
\begin{equation}
\begin{array}{c}
 \hspace{0.4cm} S_{\alpha \beta }  = \frac{{e^2 \omega}}{ \pi }\left[ {\delta _{\alpha \beta } \sum\limits_{\beta j_1 j_2 } {X_{\omega ,j_1 } X_{\omega ,j_2 } \frac{{\partial S_{\alpha \beta } }}{{\partial X_{j_1 } }}\frac{{\partial S_{\alpha \beta }^* }}{{\partial X_{j_2 } }}2\cos \left( {\varphi _{j_1 }  - \varphi _{j_2 } } \right)} } \right. \\
 \hspace{3cm}  \left. { - \sum\limits_{\gamma _1 j_1 \gamma _2 j_2 } {S_{\beta \gamma _1 }^* S_{\beta \gamma _2 } X_{\omega ,j_1 } X_{\omega ,j_2 } \frac{{\partial S_{\alpha \gamma _1 }^* }}{{\partial X_{j_1 } }}\frac{{\partial S_{\alpha \gamma _2 } }}{{\partial X_{j_2 } }}2\cos \left( {\varphi _{j_1 }  - \varphi _{j_2 } } \right)} }
 \right].
 \\
 \end{array}
 \end{equation}
Here, we remark that capturing the sensitivity of the quantum
levels and scattering matrix to general perturbations is a
complicated problem entailing the treatment of a
time-dependent Hamiltonian. In the limit of an adiabatic
pump however, i.e., assuming that the scattering properties follow the
time-dependent potentials instantaneously, it is sufficient to
expand the time-dependent scattering matrix to first order in the
frequency. In addition, the amplitudes $X_{\omega ,j} $ are chosen small
with respect to their residual values (e.g., the amplitude of $X_2$ is smaller than $\Theta$) such that
 only the terms linear in $X_{\omega ,j} $ are relevant in an expansion of
the scattering matrix, which leads to a (bi)linear response in the
amplitudes\cite{Ref2, Ref3}. The pumped current and the noise relate to
the parametric derivatives of the scattering matrix of the
system\cite{Ref2, Ref3}.

For the consideration of the time-reversal symmetry (TRS) in this
system we remark the following. The (non-diffusive) scattering
region consists of
 non-collinear localized magnetic moments (that build the DWs) which
 reverse sign upon time-reversal. This operation leads
  in general to  different scattering and pump properties \cite{trs}.
  In our particular case however,
 upon time-reversal, the spin polarizations of the states left and right to
 the DWs is reversed and outgoing
waves turn into incoming waves and vice versa. In total, even though
the individual
 wave functions and the scattering
region are modified upon a time-reversal, we observe no physical
effect of this operation.
 The situation would be different, if for example the wire to the right side of the DWs were
 paramagnetic. An elaboration on this point is given below.

  Here, we start by specifying the single
particle states and calculate with those the T matrices. The
time-reversal of the carrier wave functions (2) can be obtained as
\begin{equation}
 \begin{array}{l}
 \begin{array}{*{20}c}
   {\psi _k \left( z \right) =
   \left[\left( {e^{ - ikz}  + R^* e^{ikz} } \right)\left|  \downarrow  \right\rangle  + R_f^* e^{\kappa z} \left|  \uparrow  \right\rangle
    \right] e^{iEt} ,} & {z <  - L,}  \\
\end{array} \\
 \begin{array}{*{20}c}
   {\psi _k \left( z \right) = \left[\left( {A^* e^{\kappa z}  + B^* e^{ - \kappa z} } \right)\left|  \downarrow  \right\rangle  + \left( {C^* e^{ - ikz}  + D^* e^{ikz} } \right)\left|  \uparrow  \right\rangle \right] e^{iEt},} & {\left| z \right| < L,}  \\
\end{array} \\
 \begin{array}{*{20}c}
   {\psi _k \left( z \right) = \left[T^* e^{ - ikz} \left|  \downarrow  \right\rangle  + T_f^* e^{ - \kappa z} \left|  \uparrow  \right\rangle \right] e^{iEt},} & {z > L.}  \\
\end{array} \\
 \end{array}
 %\right.} \right) \times e^{{{iEt} \mathord{\left/
 %{\vphantom {{iEt} \hbar }} \right.
 %\kern-\nulldelimiterspace} \hbar }} .
\end{equation}
Spin-up electrons tunnel through the double-domain-wall structure by
spin-flip-assisted transmission. When the time is reversed, the
angular momentum of the electrons, here  the spin, is reversed. As
time flows backwards, spin-down electrons tunnel backward through
the double-domain-wall structure by spin-flip-assisted transmission.
The time-reversed scattering matrix is the Hermitian conjugate of
the original one with the transmission probability exactly
identical. The nature of DWs is that of an angular momentum. Time
reversal reverses the sign of the localized moments,  forming thus
an anti-double-domain-wall structure with a spin-down well between
the two domain walls. Outside the domain wall structure, the
time-reversed magnetic nanowire  favors then a transport of
spin-down electrons. In the adiabatic limit, the applied external
magnetic field is effectively incorporated in the behavior of the
DWs. Or it is equivalent to say that the effect of the time-reversal
transformation reverses the direction of the magnetic field and that
reverses the magnetic configuration. To contrast this situation with
what is established in the literature we recall  that  in
Ref.[\onlinecite{Ref2}] Brouwer considered  a chaotic quantum dot
with the conclusion  that the distribution of the pumped current for
systems with TRS and those without TRS are remarkably different. In
our consideration the strength of the particle flow is conserved
through tunneling, which is required by the continuity equation for
the Schr\"{o}dinger equation. The time reversed tunneling would
generate exactly the same pumped current with the electron spin
reversed. In our case, the distribution of the pumped current is not
a defined quantity or is a constant.

Under the assumption that the scattering properties follow the
time-dependent potentials instantaneously, calculating the
parametric derivative of the scattering matrix is a numerical issue
as demonstrated in the literature\cite{Ref3, Ref4} including Brouwer's
seminal approach\cite{Ref2}. In our work, the stationary
transmission and reflection coefficients were analytical obtained
along with
 their parametric derivatives. The derivative of the
scattering matrix can be inferred  from the stationary Hamiltonian:
an infinitesimal change in the external parameters (here, the
magnetization field and the Fermi energy) alters the Hamiltonian and
hence the scattering matrix. Accordingly, in the adiabatic limit, we
can find the derivative without introducing a time-dependent
Hamiltonian. In the linear response situation, all the sensitivity
of the scattering matrix to the perturbations lies in its parametric
derivatives.

Considering an adiabatic pump, the modulation frequency $\omega$ is
assumed to be extremely small relative to the interaction time of
the system. The current and noise vary linearly with $\omega$,
  which is within the theory of an adiabatic pump\cite{Ref3} and
  demonstrated in experiment\cite{Ref1}.
  %Thus in our approach, the frequency is universally the same
%  in all calculations and presents
%  in the units of the current, noise, and heat flow.
  The first term in Eq. (7) reflects the strength of the energy flow
while the second one  describes the effect of correlations between
(quasi)-particles. A convenient measure of the correlation is
$F=\frac{{\hbar  \omega  }}{{4e^2
}}\frac{S_{\alpha\beta}}{H_\alpha}$, which is the ratio of the
dimensionless strength of the shot noise $S_{\alpha\beta}$ and the
heat flow. The pump current and the noise experience sinusoidal and
cosinusoidal variations as functions of the phase difference $\Delta
\varphi  = \varphi _1  - \varphi _2 $ respectively, as is evident in
Eqs. (5) and (7). The relative noise $F$ does not vary with $\Delta
\varphi$.
\section{Numerical results and interpretations}
In a magnetic nanowire with double sharp magnetic DWs, quantum
interference results in the formation of spin-split quasistationary
states localized mainly between the domain walls\cite{Ref24}.
Consequently, the DWs conductance exhibits typical resonant
tunneling behaviour. The width of the resonance peaks is related to
$\tau$, the lifetime of the quasistationary spin quantum well
states. $\tau$ is determined by the spin-mixing
 due to the spin noncollinearity
at the DWs and can be quantified by the spin-mixing parameter
$\Delta  \equiv 4m\Theta\delta /\hbar ^2  $. Decreasing $\Delta$, the
lifetime of the localized spin quantum-well states increases and the
conductance resonance peaks become correspondingly narrower. By
oscillating the gate potential and the magnetic field strength out
of phase at zero bias, a quantum pump is realized as the electronic
system gains energy from the oscillating scatterer. Absorption of an
energy quantum $\hbar \omega $ leads to creation of a nonequilibrium
quasi-electron-hole pair. If they are scattered into different
leads, their motion generates current.

In the adiabatic quantum pump based on domain walls in a magnetic
nanowire considered here, the two parameters are the Fermi energy
and the magnetization strength, which excite the pump. The Fermi
energy can be modulated by an ac gate voltage and the magnetization
strength can be modulated by an ac magnetic field. Analogously to
 the adiabatic quantum
electron pump theoretically proposed in Ref. \onlinecite{Ref2} and experimentally
realized in the quantum dot system\cite{Ref1}, in our
case the oscillating
magnetization acts as an oscillating energy potential  (cf. Fig.1).
 The two oscillating parameters we view
 as two out-of-phase gating potentials, one from the ac
electric modulation, and the other from the ac magnetic modulation;
on the other hand the quantum dot pump system experiences
 two oscillating gates. Electrons absorb or emit an energy
quantum from the oscillating mesoscopic scatterer and those pumped
to different directions contribute to the current. The external
electric gate voltage oscillating at the spin-split quasistationary
states excites pumping current and gives rise to the resonance-type
peaks in the current. Within  the two-gate picture, the DW quantum pump system can
be viewed semiclassically as in the turnstile quantum dot---a
classical analog of quantum pumping\cite{Ref32}.

 Fig. 2 shows  the  tunneling current without external
modulations as well as the current adiabatically pumped  by the
modulations (4)
 as functions of the Fermi energy.
In Fig. 2 (b), the phase difference $\Delta \varphi $ is set to be $
- \pi /2$ to produce maximal pump current and to demonstrate the
pumping properties prominently. The conductance of the double DW
structure possesses sharp resonances when the Fermi energy matches
the spin quantum-well states as seen in Fig 2 (a) and
Ref.\onlinecite{Ref24}. Similarly, in the circumstance of the
adiabatic pump, the electron can only tunnel through the spin-split
levels between the two domains. Thus, the pump current, shot noise
and heat flow show peak structures when the Fermi energy matches the
spin quantum-well states (see also Fig. 3).

Comparing panels (a) and (b) in Fig. 2 we find that the peak height
(PH) of the pumped current decreases dramatically as the width of
the DWs is increased. For narrower DWs, the peak width (PW) of the
pumped current becomes more comparable with that of the
linear-response conductance, while for wider DWs the former is much
smaller than the latter. To further reveal the trend, the PH and the
PW at half peak height of the pumped current versus $\delta$ are
given in Figs. 4 and 5 respectively for the second quasistationary
level counting from the spin quantum-well bottom. The peak height
decreases exponentially with the increase of $\delta$. In contrast
to the linear-response process, the pumping process depends strongly
on the lifetime of the quasistationary, spin-well states. When the
period of the pump $\tau _p  = 2\pi /\omega $ exceeds the lifetime
of the quasistationary states, the pumped current would be greatly
suppressed. Here, we consider the limit of small modulation
frequencies, i.e. an adiabatic pump\cite{Ref3} meaning that  $\tau
_p$ is relatively large. Therefore, the strength of the pumped
current decreases exponentially with the widening of the DW, i.e.,
the decrease of the lifetime of the quasistationary states.

In the adiabatic limit, i.e. when the frequency of the potential
modulation is small compared to the characteristic times for
traversal and reflection of electrons, the photon side band $E_F
\approx E_r  \pm \hbar \omega $ broadens the current peak instead of
forming new peaks as in Ref.\onlinecite{Ref5}. In principle, without
the external modulations the resonant tunnelling current is
broadened due to spin mixing at the DWs. If the spin-mixing
amplitude is finite (i.e., $\Delta   \ne 0$), spin-up carriers
transverse resonantly the DWs and the conductance peaks are
broadened by the spin-mixing mechanism. For very narrow DWs the
residual broadening can then be attributed to the side-band
formation.

In the theory of an adiabatic quantum pump, the nearest sidebands
corresponding to particles which have gained or lost a modulation
quantum $\hbar \omega $ contribute to the pumped current. Hence, the
pumped current peak is broadened by $2 \hbar \omega $ at the
resonant levels. The width of peaks in the linear-response
conductance due to spin mixing at the DWs is enhanced  for wider DWs
(cf. Fig. 5). In comparison, the broadening of the pumped current
peaks  caused by  the sideband formation is less pronounced for
varying DW extensions.  For very sharp DWs however the effect of the
side-band formations might be as large as the contribution from the
DWs spin mixing.

To demonstrate the quantum pump noise properties in this particular
structure, we present the numerically obtained pump current $I$,
shot noise $S$, heat flow $H$, and relative pump noise $F$ as
functions of the Fermi energy in Fig. 3. The properties for
different DW distance $L$ are compared. The pump current, shot noise
and heat flow show peak structure when the Fermi energy matches with
the resonant energy levels of the structure. The peak position
changes accordingly as the resonant levels shift to lower energy
when we increase the DW distance $L$. The ratio between the pump
shot noise and the heat flow $F$ reflects the strength of the
quantum correlation of the system. $F$ shows bifurcated peaks with
the intermediate minima at the resonant Fermi energies.
%The sub-peak distance approximates the peak width.

The pump noise properties can be interpreted as follows. At the resonant
Fermi energy, transport processes and thus the electron-electron
correlation achieve maximal strength. The nonequilibrium
quasi-electrons and holes created by the oscillating scatterer move
in different directions to generate a net dc current. Therefore, the
antibunching correlation between electrons and holes always exerts a
negative contribution to the shot noise regardless of the direction
of the dc current, which gives rise to the minimum valley in the
relative noise. At the edges of the resonant state, there is a
slight heat flow in the dissipation regime without actual particle
motion. In this region, the heat flow dominates the quasiparticle
correlation giving rise to full heat flow noise with $F=1$. In the
intervals between resonant levels, heat flow and quasiparticle
correlation both approach a background level inducing zero pump shot
noise.

\section{Conclusions}
In summary, a quantum pump device involving domain walls in a
magnetic nanowire is investigated. For two independent adiabatically
modulated  parameters of this device a finite net charge current is
transported. The quantum pumping mechanism
resembles  the quantum pump based on two-oscillating-gate quantum dot
 and is to some extent analogous to the  classical
turnstile picture. The pumping
current, heat flow, and shot noise demonstrate peak structures at
the spin-split quasistationary levels in the spin-quantum well
formed by  the domain walls.
 The strength of the pump current decreases
exponentially with the decrease of the quasistationary-state
lifetime. The latter is governed by the width of the domain wall. The sideband
formation during the pumping process is sizable particularly for
narrow domain wall for which level broadening due to spin-mixing is relatively small.
The correlation between
quasielectrons and quasiholes shows antibunching behavior as
they move in opposite  directions. This is concluded  from the ratio
between the pump shot noise and the heat flow.

\clearpage

\begin{figure}[t]
 \includegraphics[width=10cm]{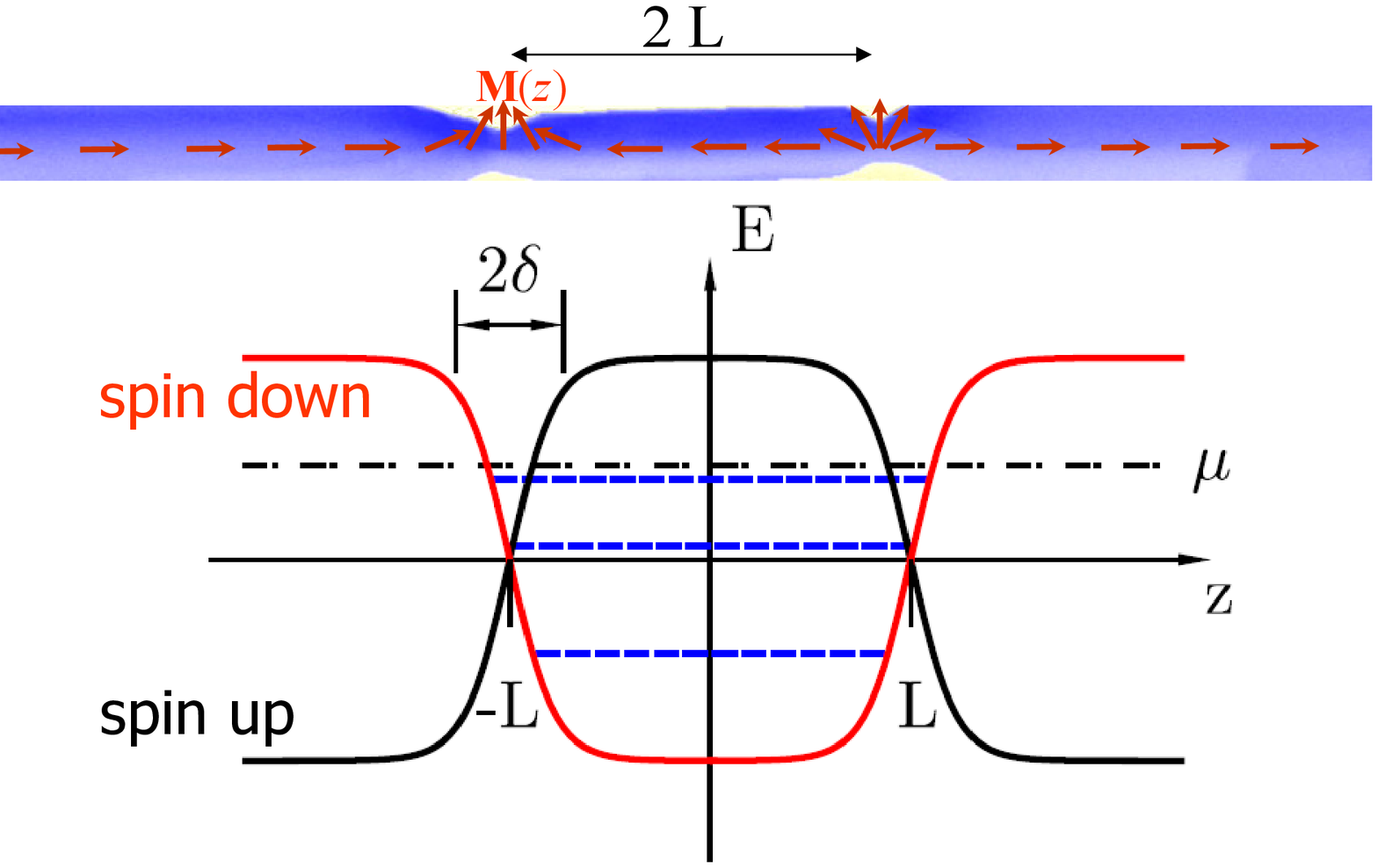}
\caption{(upper panel) Schematics of the variation of the magnetization $\mathbf{M}$ in a
magnetic quantum wire with two domain walls
(DWs). The easy plane  is chosen as the $x$-$z$ plane. DWs are separated by a distance $2L$.
 (lower panel)  the effective potential  profile experienced by   spin-up
and spin-down electrons.  Quasi-localized energy levels are marked by
 dashed lines. $2 \delta$ is
the DW width and $\mu$ is the chemical potential.
}
\end{figure}

\begin{figure}[h]
 \includegraphics[width=10cm]{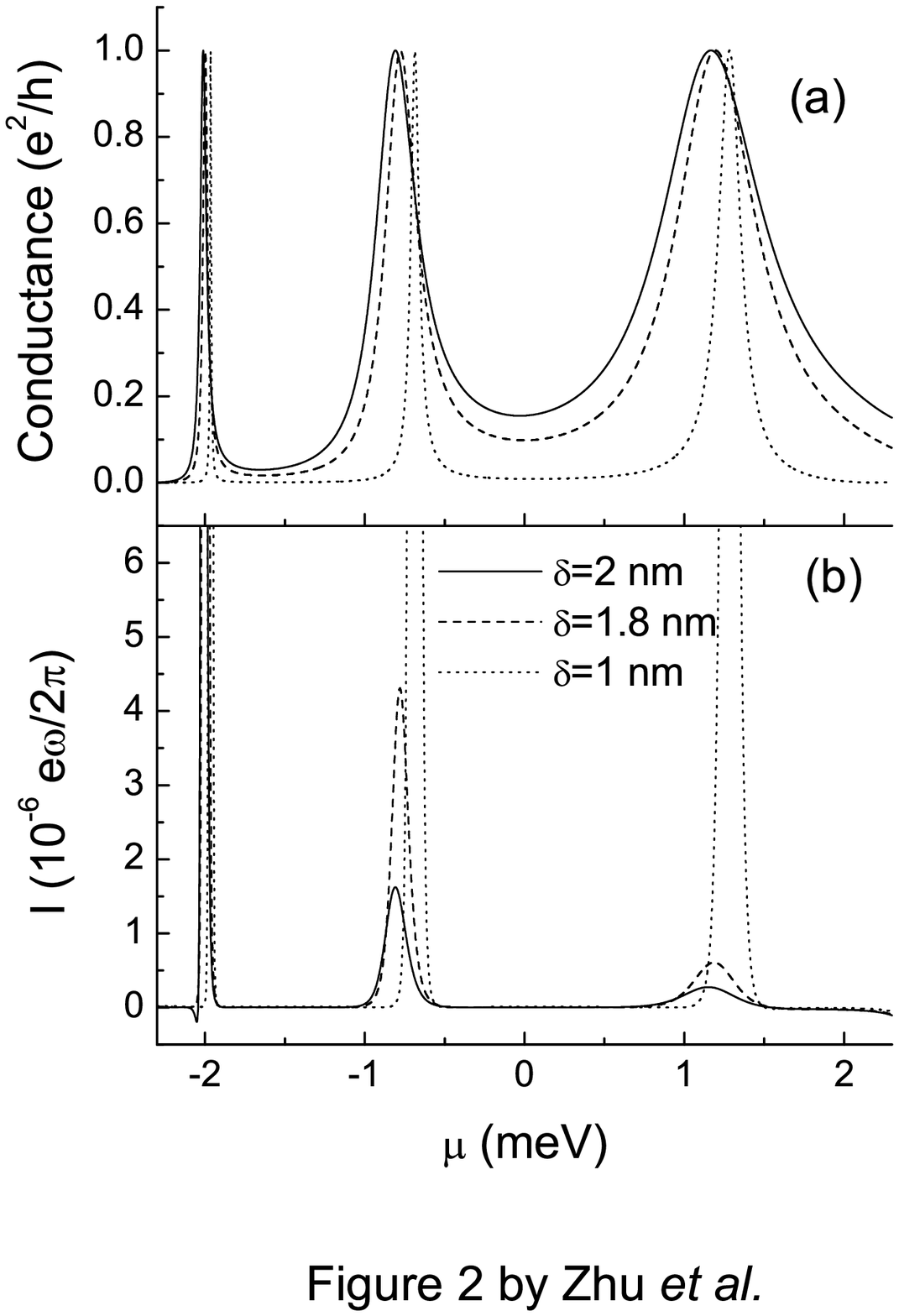}
\caption{Linear-response conductance (a) and pump current $I$ (b) as
functions of the Fermi energy $E_F$ for different DW width $\delta$.
The parameters are chosen as follows: The coupling of the carriers
 to the DWs is $\Theta \equiv JM = 2.4$ meV; the amplitudes  of the
 external perturbations  are $\Theta _\omega =X_{1,\omega}=0.1$
 meV $ = \mu_\omega = X_{2,\omega}$; the distance between the DWs is  $L = 15$ nm.
In panel (b), the parameters
are set in the weak pumping regime.}
\end{figure}

\begin{figure}[h]
 \includegraphics[width=10cm]{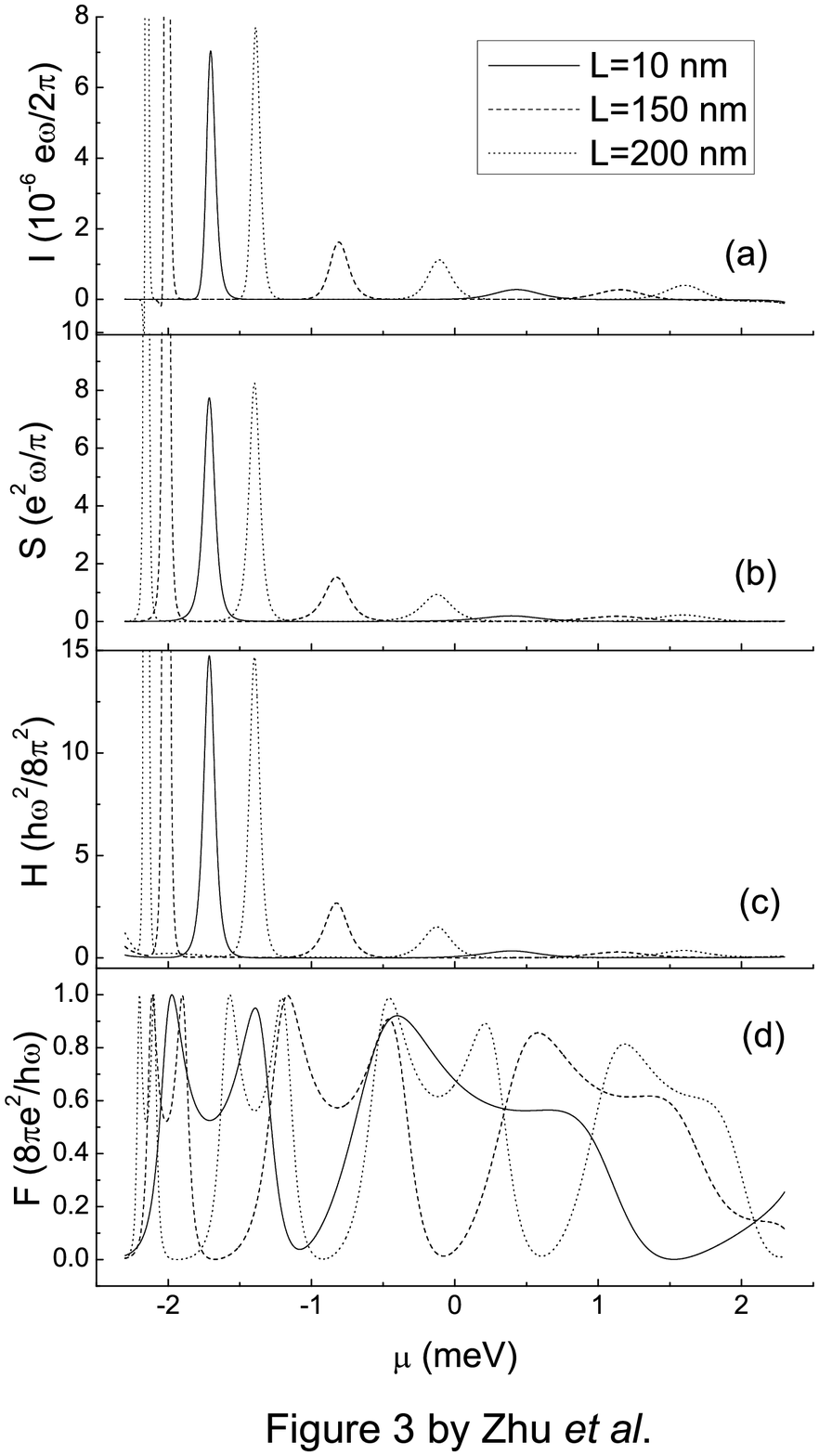}
\caption{Pump current $I$ (a), pump noise $S$ (b), heat flow $H$
(c), and relative noise $F$ (d) as functions of the Fermi energy
$E_F$ for different DW distance $L$. DWs width is $\delta  = 2$ nm and the other
parameters are the same to Fig. 1 (b).}
\end{figure}

\begin{figure}[h]
 \includegraphics[width=10cm]{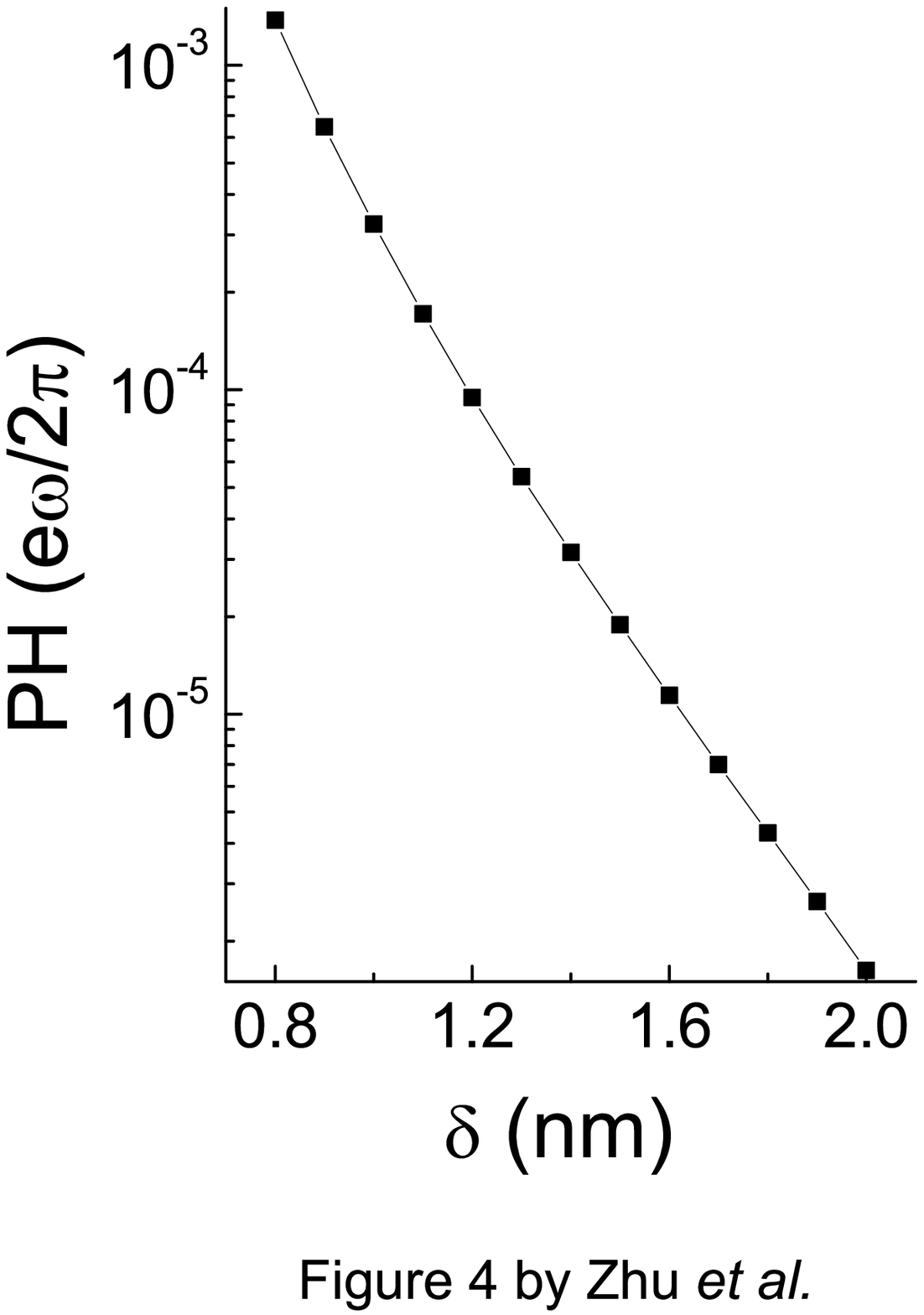}
\caption{The peak height (PH) of the pumped current as a function of
the DW width $\delta$ for the second quasistationary level counting
from the spin well bottom. The parameters are the same to Fig. 1
(b).}
\end{figure}

\begin{figure}[h]
 \includegraphics[width=10cm]{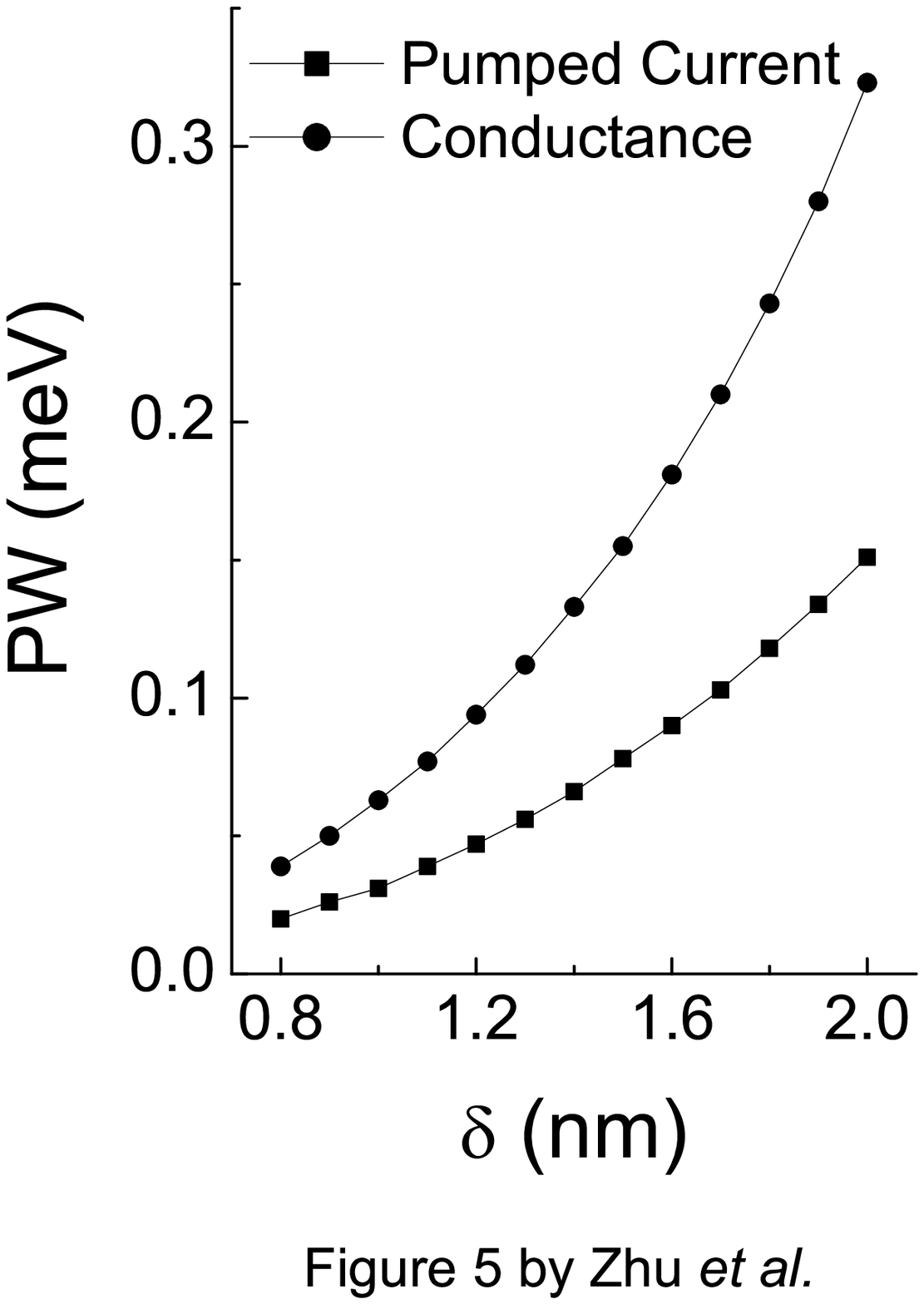}
\caption{The peak width (PW) at half peak height of the
linear-response conductance (dot) and the pumped current (square)
respectively as a function of the DW width $\delta$ for the second
quasistationary level counting from the spin well bottom. The
parameters are the same to Fig. 1 (b).}
\end{figure}


\begin{references}
\bibitem{Ref30} When the pumping mechanism is taken into account,
the scattering matrix elements vary with the oscillating parameters.
A scattered electron can absorb or emit one energy quantum $\hbar
\omega $ before it leaves the scattering region. This does not
contradict our single-channel picture.
\bibitem{Ref31} In most quantum transport problems
the transmission coefficients seen from the two sides of the scattering region differ at
least in phase.  Structural configuration
symmetry imposes however the relations $t'=t$ and $r'=r$.
The structure of the double-DW nanowire system is symmetric. The
band edge does not vary in the transmission direction, which means
that the band profile seen from the left is exactly the same with
that seen from the right (cf. Fig.1).  Having derived
analytically  $t'$ and $r'$ from spatially inverted wave
functions we confirmed explicitly  the above statement.
\end{references}
\end{document}